\documentclass[final,3p,times]{elsarticle}

\usepackage{amsmath}

\usepackage{amssymb}

\biboptions{sort&compress}

\newcommand{\ba}{\begin{eqnarray}}
\newcommand{\ea}[1]{\label{#1} \end{eqnarray} }

\begin{document}

\begin{frontmatter}



\title{\bf \Large Wealth distribution in modern societies: \\
 collected data and a master equation approach}


\author[label1]{Istv\'an Gere}, 
\author[label1]{Szabolcs Kelemen} 
\author[label2,label3]{G\'eza T\'oth} 
\author[label4,label5]{Tam\'as S. Bir\'o}
\author[label1]{and Zolt\'an N\'eda}

\address[label1]{Babe\c{s}-Bolyai University, Dept. of Physics, Cluj-Napoca, Romania}
\address[label2]{Central Statistical Office of Hungary, Budapest, Hungary}
\address[label3]{University of Miskolc, Department of Economics, Miskolc, Hungary}
\address[label4]{Wigner Research Center for Physics, Budapest, Hungary\footnote{Part of the Eotvos Roland Research Network ELKH}}
\address[label5]{ Complexity Science Hub, Vienna, Austria}

\begin{abstract}
A mean-field like stochastic evolution equation with growth and reset terms (LGGR model) is used to model wealth distribution in modern societies. The stationary solution of the model leads to an analytical form for the density function that is successful in describing the observed data for all wealth categories. In the limit of high wealth values the proposed density function has the accepted Tsallis-Pareto shape. Our results are in agreement with the predictions of an earlier approach based on a mean-field like wealth exchange process. 
\end{abstract}




\end{frontmatter}







\section{Introduction}
Since the last decade of the $19^{th}$ Century, when the pioneering studies on inequalities in socio-economic systems were performed by Vilfredo Pareto \cite{pareto}, the distribution of wealth and income have been intensively studied \cite{pikety}. According to Pareto's well-known result the richest end of the cumulative distribution for wealth and income in a given society follows a power-law function characterized by the so-called Pareto exponent. Probably this was the first encounter of the scientific community with non-Gaussian, scale-free distributions in complex socio-economic systems. It was lately recognized that the low and middle range of the wealth and income distributions, follow a different trend, which was many times approximated by a Boltzmann-Gibbs type  exponential distribution \cite{yakovenko,Cui,Cardoso,Jongsoon,Lim}.

Nowadays, many electronic databases containing a large amount of data on income and wealth in different countries, are accessible for researchers \cite{datasets,Cao,solomon}. Such data can be a goldmine for those, who wish to explore social inequalities, universalities or dynamics in the distribution of socio-economic proxies. The data for both quantities (wealth and income) can be studied on the level of individuals or groups (families, companies, settlements, ...) \cite{Wolf,Clementi,dunford}, with an immediate influence on the shape of the obtained distribution. Although, there might be correlation between the income and wealth of the individuals \cite{similarities,yakovenko09}, there is no direct connections between these two economical measures.   A universal and striking similarity between wealth and income is that for both of them Pareto's law apply: the tail is power-law like. Nevertheless in the limit of small and middle wealth/income values the distribution function for these two quantities can have a different shape \cite{yakovenko,similarities}. 
Due to the fact that the income of individuals in a society is directly derivable from tax data, there are excellent exhaustive databases for this socio-economic measure making the investigations more easy \cite{noemi}.  Sampling for wealth distribution is more peculiar and less accurate than it is in case of the income. The data that is available for wealth is mostly based on using some indirect measures (proxies), estimations and annual surveys. A relevant difference relative to income is also the fact that wealth can be negative, meaning debts. In such a view the distribution function of wealth should be more complex, and should not be limited to the  $[0,\infty]$ interval.  

\par
In the last few decades many theoretical work were  done both by economists
\cite{levy,jones,sorger} and physicists \cite{yakovenko, Cui, Cardoso, Jongsoon, Lim, Clementi, Bouchaud, Chatterjee, Oliveira, ZNeda_gambler, ZNeda_family, ZNeda1, model_pareto, Ciesla}  to understand the measured income and wealth distributions. 
Based on how these models target the dynamics of the relevant economic quantity (wealth or income)  the majority of them can be grouped in a few relevant classes \cite{yakovenko}. The first type models from the literature are based on simple analogies with thermodynamic systems. The relevant distributions are usually derived either from maximization of the Shannon-Gibbs entropy under different conditions or by the generalization of the concept of entropy \cite{Cui,Lim,Tsallis}.  The second type models describe the relevant economic phenomenon as stochastic exchange processes based on predefined dynamical rules. Such an approach can be either an agent-based computation \cite{comp} or a mean-field type analytical model. 
As an example for such a process, that one should mention here, is given by the multiplicative growth and exchange model, elaborated by J.P. Bouchaud and M. Mezard \cite{Bouchaud}, that will be discussed in more detail later.   
Approaching the dynamics of income and wealth based on master equations with average growth, decrease or reset rates \cite{Cardoso, ZNeda1} is also a common modeling paradigm. Apart of these two main categories there are also a large variety of models steping over the mean-field type approximations and considering exchange processes on lattice \cite{Cui, Jongsoon} or random networks \cite{Clementi, ZNeda_gambler, ZNeda_family}.

\par
Recently, we proposed a simple model  based on mean-field like stochastic growth and reset processes for describing income distribution in modern societies \cite{ZNeda1}. The used master-equation contains growth and reset rates derived form real world data. 
The model offered an excellent description for the income distribution in all income categories \cite{ZNeda1}. 
Here we aim to show that a similar approach is successful in modeling the collected data for wealth as well. 
After our knowledge a model that is successful in describing analytically the distribution of wealth for all wealth categories is still missing. 
Therefore, as a step forward relative to the presently available wealth distribution models,  we plan to give a unified and compact analytical description for the density function of wealth distribution on the entire wealth interval. 
Our manuscript is organized in the following manner: (1) first we present the used modeling framework and apply it to the wealth dynamics in social systems, 
(2) we than derive from statistical data the relevant density functions for wealth distribution in modern societies and compare them with the analytical prediction of our model, 
(3) finally we discuss the agreement between these data and our model predictions and comment on the appropriateness of the applied stochastic growth and reset rates.

\section{The growth and reset process}
\newcommand{\be}{ \begin{equation} }
\newcommand{\ee}[1]{\label{#1} \end{equation} }

A simple local growth and global reset master equation 
(LGGR model) proved to be successful in modeling relevant statistics in several 
complex phenomena \cite{Biro-Neda}.
The evolution equation contains both local and long distance transitions: 
uni-directional local growth and reset to a given new state \cite{Biro-Neda}. 
In order to present this model let us consider an ensemble of identical elements 
that can have different  numbers of quanta  of a relevant quantity. 
An immediate example in the line of the problem considered here, 
are the individuals in a society, owning different amount of wealth.
Let us denote here by $P_n(t)$ the probability that a person has exactly $n$ quanta 
of wealth at time $t$. Normalization requires $\sum_n P_n(t)=1$.  
In case the reset is only to the state with $n=0$ quanta the dynamics of the growth and reset 
process in the space of the wealth quanta $n$ is sketched in Figure \ref{smart-reset}a. 
For this case the evolution equation for the $P_n(t)$ probabilities writes as:  

\begin{equation}
\frac{dP_n(t)}{dt}=\mu_{n-1}P_{n-1}(t)-\mu_n P_n(t)-\gamma_nP_n(t) + \delta_{n,0}\langle \gamma \rangle(t) . 
\label{master_dis}
\end{equation} 
Here we denoted the growth-rate from state $n$ to $n+1$ by $\mu_n$ and the reset rate from the state with $n$ quanta to state $n=0$ as $\gamma_n$. 
The last term on the right side is re-feeding at state $n=0$, ensuring the preservation of normalization for $P_n(t)$:
\begin{equation}
\langle \gamma \rangle(t) =\sum_{j} \gamma_j P_j(t).
\end{equation}

The dynamical process from above can be generalized to continuous quanta, $n\rightarrow x \in R$. 
Instead of the discrete $P_n(t)$ probabilities we shift to a continuous $\rho(x,t)$ probability density, 
with the normalization condition $\int_{\{x\}} \rho(x,t) dx=1$. As it is detailed in \cite{Biro-Neda}
the continuous limit of the master equation (\ref{master_dis}) becomes:

 \begin{equation}
 \frac{\partial \rho(x,t)}{\partial t}=-\frac{\partial}{\partial x} \left[ \mu(x) \rho(x,t) \right] - \gamma(x) \rho(x,t) +\langle \gamma(x) \rangle (t) \delta(x),
 \label{master_gen}.
 \end{equation}
 The re-feeding at $x=0$ and conservation of the normalization is ensured by the term with the $\delta(x)$ Dirac functional and by considering:
 \begin{equation}
 \langle \gamma(x) \rangle (t) = \int_{\{x\}} \gamma(x) \rho(x,t) dx
 \label{normal}
 \end{equation} 
 
 \newcommand{\eadx}[1]{{\rm e}^{#1} }
  In previous studies it  was proven, that the above dynamical evolution equation converges to a steady-state with the
 $\rho_s(x)$ stationary probability density \cite{Biro-Neda,BNT}
 \begin{equation}
\rho_s(x) \: = \: \frac{\mu_0 \rho_s(0)}{\mu(x)} \, \eadx{-\int_0^x\limits \frac{\gamma(u)}{\mu(u)}du},
\label{stat-distr}
 \end{equation}
 with:
 \begin{equation}
 \langle \gamma(x) \rangle = \int_{0}^{\infty} \gamma(x) \rho_s(x) dx
 \label{normalx}
 \end{equation} 

As we discussed in several recent publications, many important distributions 
that are frequently encountered in complex systems can be explained in the framework 
of the growth and reset model by properly selecting the state-dependent local 
growth rate $\mu(x)$ and reset rate $\gamma(x)$.
Among these studies our recent one of income distribution \cite{ZNeda1} 
by the LGGR model encourages us to attempt a similar approach to wealth.

\section{Wealth distribution in view of the LGGR model}

\begin{figure}[!ht]
	\centering
		\includegraphics[width=10cm]{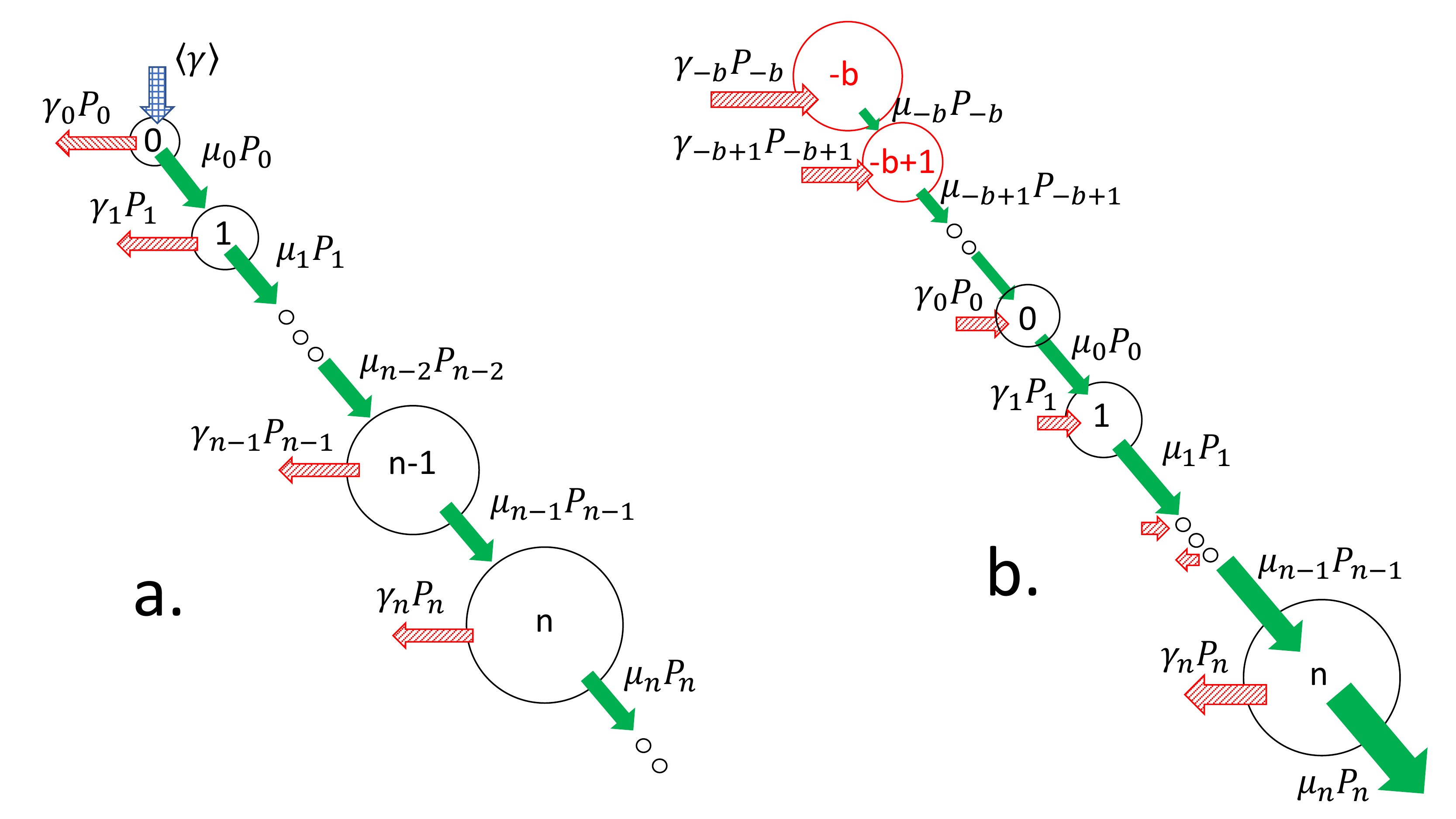}	
\caption{Schematic illustration of the growth and reset process. Figure (a.) suggests the 
general mechanism of the growth and reset process, while Figure (b.) 
shows the process with a reset rate that is negative for small wealth 
and positive for large wealth values.}
	\label{smart-reset}
\end{figure}

In order to realistically model income a linearly increasing growth rate 
\begin{equation}
\mu(x)=\sigma \:(x+b),
\label{lin-growth}
\end{equation}
and a smart-reset rate was used in \cite{ZNeda1}, allowing both 
the appearance and disappearance  of individuals in different income categories. 
For low income values the reset rate was chosen to be negative, while 
for higher income values the reset rate became positive, saturating to a finite value. 
Such a flow is illustrated schematically for the discrete probability space in 
Figure \ref{smart-reset}. The negative reset rate at low $x$ values
describes  an appearence of new individuals in the system with that income, 
the positive reset rate on the other hand means disappearance of individuals 
from the income category at $x$. 
The investigation of an exhaustive ten year social security data in Cluj county (Romania) 
confirmed the linearly increasing growth rate with $b=0$, and was supporting a kernel 
function for the reset rate in the form:
\begin{equation}
\gamma(x)=K-\frac{r}{x+q}, \, \, (K,b,q \in \mathbb{R}_+).
\label{resetrate}
\end{equation} 
Using such an approach the growth and reset model yields  
a Beta Prime stationary distribution for income, in excellent agreement with 
recent statistical data \cite{ZNeda1}.

The main difference 
between the distributions of income and wealth is that the total wealth of the 
individuals can be negative,  i.e. debts.  In such a context the wealth distribution 
function is
defined usually on an interval $[-b,\infty]$, while income distribution is defined 
on the $[0,\infty]$ interval. The value $b$ characterizes the maximum amount of debts 
that are accepted for the individuals  by the financing system. 
This is the amount of debt that is considered to be reimbursable. 

Similar to income, one can admit that the linearly increasing growth rate is a 
reasonable assumption also for wealth, in agreement with the Matthew's principle:
"The rich gets richer". 
The average increase in wealth over a fixed time is usually not by given amount 
but rather by a given percentage of the already existing wealth.  
In this sense the linear growth rate expressed in Equation (\ref{lin-growth}) 
should be a valid approximation for wealth, too. This growth rate is positive for 
the whole $[-b,\infty]$ interval. For $x>-b$ it allows a growth in wealth, 
therefore the reimbursement of the debts. For $x<-b$ the growth rate becomes negative, 
and therefore accumulated debts cannot be reimbursed.  

For the reset process a smart-reset rate similar to the one used for income, 
eq. (\ref{resetrate}), is a reasonable approximation for the dynamics in wealth. 
The growth process starts mainly at negative or low wealth values, 
while individuals will leave the society with higher wealth values. 
Therefore, similar to the income, the reset-rate should be negative at 
negative wealth values and saturate at a positive value for high wealth values.   
Since the negative wealth means debt, this reset rate indicates that the growth 
usually starts either from debts accumulated by loans or losses in transactions 
or from an initial low wealth value.  
Among other possibilities this includes new individuals appearing in the society 
or wealthy individuals resetting their wealth by unsuccessful transactions. 
Again, in order to prevent debts below the $b$ value, one has to choose in 
eq. (\ref{resetrate}) $q=b$.

According to the above arguments, for a simple growth and reset master equation 
approach to wealth dynamics one could use the following kernel functions for 
the growth and reset rates
\begin{eqnarray}
\mu(x)=\sigma \: (x+b) \nonumber \\
\gamma(x)=\sigma\left(k-\frac{\alpha}{x+b}\right),
\label{rates}
\end{eqnarray}
with $\alpha, k, b \in \mathbb{R}_+$.
  
In this approach it is easy to show that the normalized $\rho_s$ stationary 
distribution defined on the $[-b, \infty]$ interval becomes:
 \begin{equation}
 \rho_s(x)=\frac{\alpha ^k}{\Gamma(k)} e^{ -\frac{\alpha}{b+x}} \, (b+x)^{-1-k}.
 \label{statdistr}
 \end{equation}  
 
The reset and growth rates defined by the equations (\ref{rates}) and 
the above stationary distribution (\ref{statdistr}) ensures the conservation of 
the total number of actors ($N_{tot}$) and the total amount of wealth ($W_{tot}$) 
in the system. One can derive this directly by using the master equation (\ref{master_gen})
in the stationary limit, or by inspecting the following integrals:
 
 \begin{eqnarray}
\Delta N_{tot}&\propto &\langle \gamma(x) \rangle=\int_{-b}^{\infty} \gamma(x) \: \rho_s(x) dx=\int_{-b}^{\infty} \sigma \left( k-\frac{\alpha}{x+b}\right) \frac{\alpha ^k}{\Gamma(k)} e^{ -\frac{\alpha}{b+x}} \, (b+x)^{-1-k} dx=0,\nonumber \\
\Delta W_{tot} &\propto & \int_{-b}^{\infty} [\mu(x)-x \gamma(x)] \: \rho_s(x) \:dx = \nonumber \\
&=& \int_{-b}^{\infty} \sigma \left[(x+b)-k\,x+\frac{\alpha \,x}{x+b}\right] \frac{\alpha ^k}{\Gamma(k)} e^{ -\frac{\alpha}{b+x}} \, (b+x)^{-1-k} dx=0
\end{eqnarray}
 
The stationary distribution function (\ref{statdistr}) leads to the average wealth value,
 \begin{equation}
\langle x \rangle_s =\int_{-b}^{\infty} x \, \rho_s(x)\, dx=\frac{\alpha}{k-1}-b = a\: b
 \end{equation}
with 
 \begin{equation}
 a=\frac{\alpha}{b(k-1)}-1.
 \end{equation}
 
It is easy to show that the distribution function for the wealth normalized by the 
average wealth, $w=x/\langle x \rangle$, writes as:
\begin{equation}
\rho(w)=\frac{a\, (a+1)^k\:(k-1)^k}{\Gamma(k)}  e^{ -\frac{(a+1)(k-1)}{1+a \, w}} \, (1+ a\, w)^{-1-k}.
\label{df}
\end{equation}
Here $w\in [-1/a, \infty]$, and the above distribution function is normalized on 
this interval.

\section{Connection to the wealth distribution function proposed by Bouchaud and Mezard}

Bouchaud and Mezard  considered a  Langevin type equation with stochastic multiplicative 
growth and exchange terms to model the wealth distribution
in a closed society \cite{Bouchaud}. 
In their approach only positive wealth values were allowed and the time-evolution 
of the  wealth $W_i$ of the individual $i$, is approximated as:
 \begin{equation}
 \frac{dW_i(t)}{dt}=\eta_i(t)\, W_i(t)+\sum _{j\ne i} [J_{ij} W_j(t) -J_{ji} W_i(t)]\: .
 \end{equation}
 
The first term on the right hand-side describes multiplicative growth governed by the 
noise term, $\eta_i$, that is assumed to have a normal distribution with 
mean $\langle \eta \rangle$ and variance $2 \Theta$. The equation is invariant under 
a scale transformation $W_i \rightarrow \kappa W_i$, 
($\kappa \in \mathbb{R}_+$). 
In the mean-field limit  ($J_{ij}=J/N$) and for wealth values normalized to the 
mean ($w_i=W_i/\langle W \rangle$) 
the above evolution equation leads to an analytically solvable Fokker-Planck equation,
that has the equilibrium solution \cite{Bouchaud}:

\begin{equation}
\rho_{BM}=\frac{g^g}{\Gamma(g)} \, e^{-\frac{g}{w}} w^{-(2+g)} .
\label{pdfBM}
\end{equation}
with $g=J/\Theta$. 
The form of this distribution function is rather similar to the more general 
distribution function (\ref{df}) derived in the previous section from the LGGR model.
The obvious difference is that $\rho_{BM}$ is defined only for $\omega \in [0, \infty]$ 
and does not incorporate the possibility of having debts. 
While the distribution function proposed by Bouchaud and Mezard has only one 
free fitting parameter, the generalized version given by equation (\ref{df})  
has two adjustable parameters, allowing for more freedom in fitting the 
observed real world distributions. 

At this point it is interesting to note that two very different mean-field 
type approaches for the wealth dynamics leads to a similar form of 
the stationary distribution function. 
The approach considered by Bouchaud and Mezard considered mean-field like exchange, 
and allowed a diffusion governed by a multiplicative noise. In contrast, 
our approach in the present study is based on a simple growth and reset 
master-equation, coarse-graining over the diffusion and exchange terms and 
incorporating these in the phenomenological growth and reset processes.

\section{Comparison with data  }
 
\begin{figure}[!h]
    \centering
    {\includegraphics[width=0.48\textwidth]{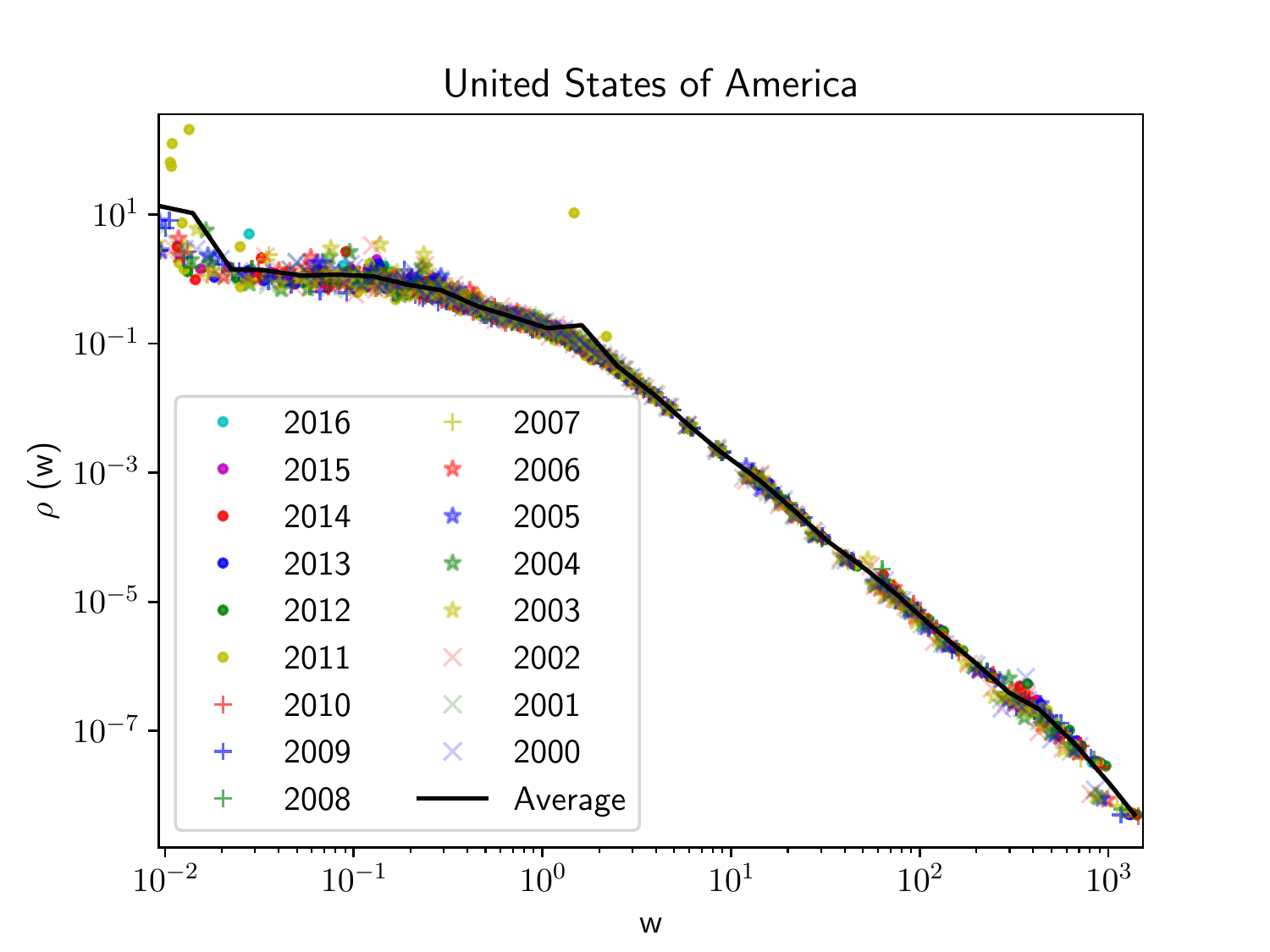}} 
    {\includegraphics[width=0.48\textwidth]{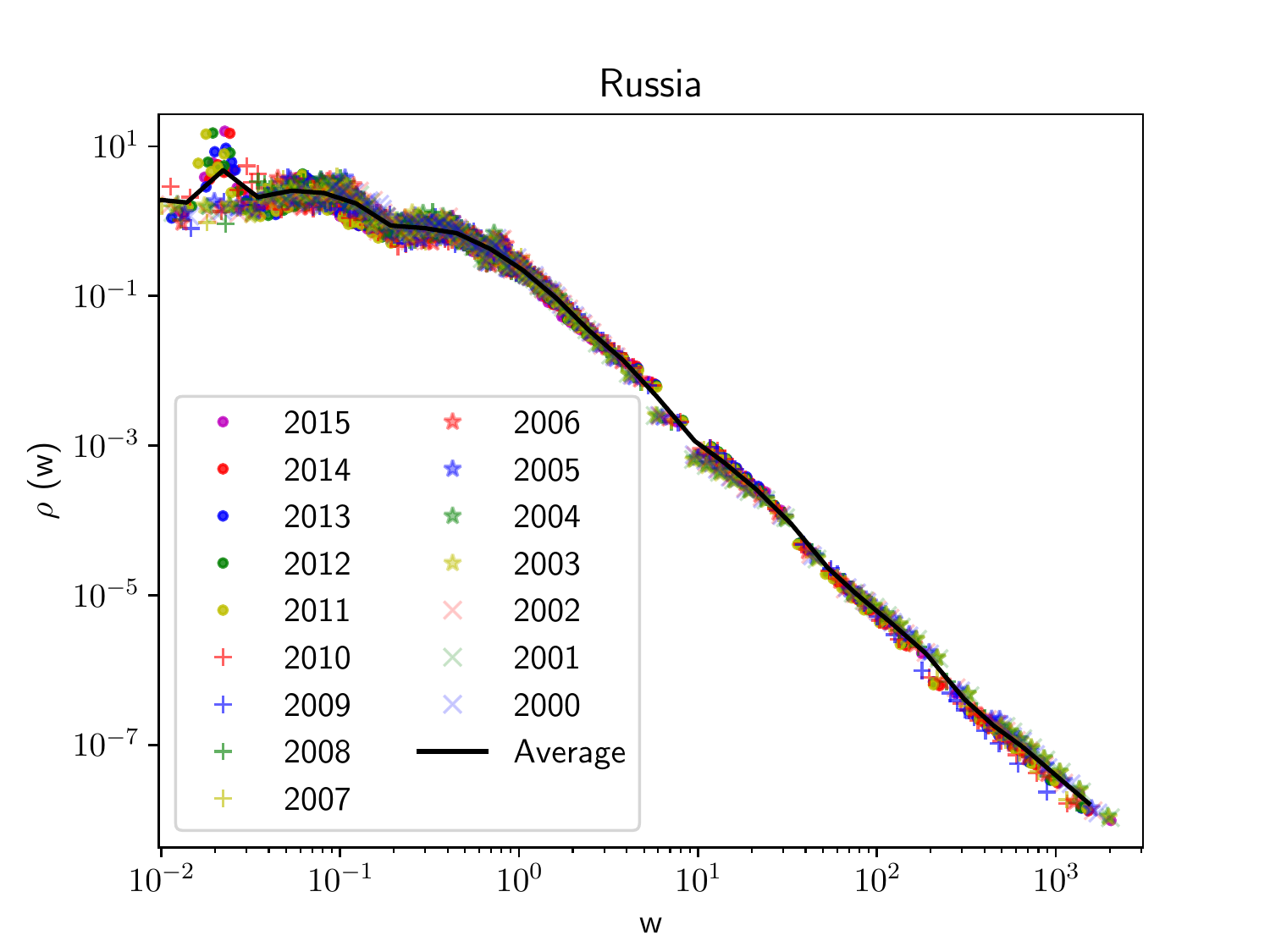}}     
\caption{Probability density function of the  normalized (rescaled) wealth, 
$w=W/\langle W \rangle$, for the $W>0$ region using log-log scales. 
Results for USA and Russia  for the years indicated indicated in the legends. 
The continuous black curve illustrates the average of the density functions for 
the considered years.}
    \label{collapse_poz}
\end{figure}

\begin{figure}[!h]
    \centering
    {\includegraphics[width=0.4\textwidth]{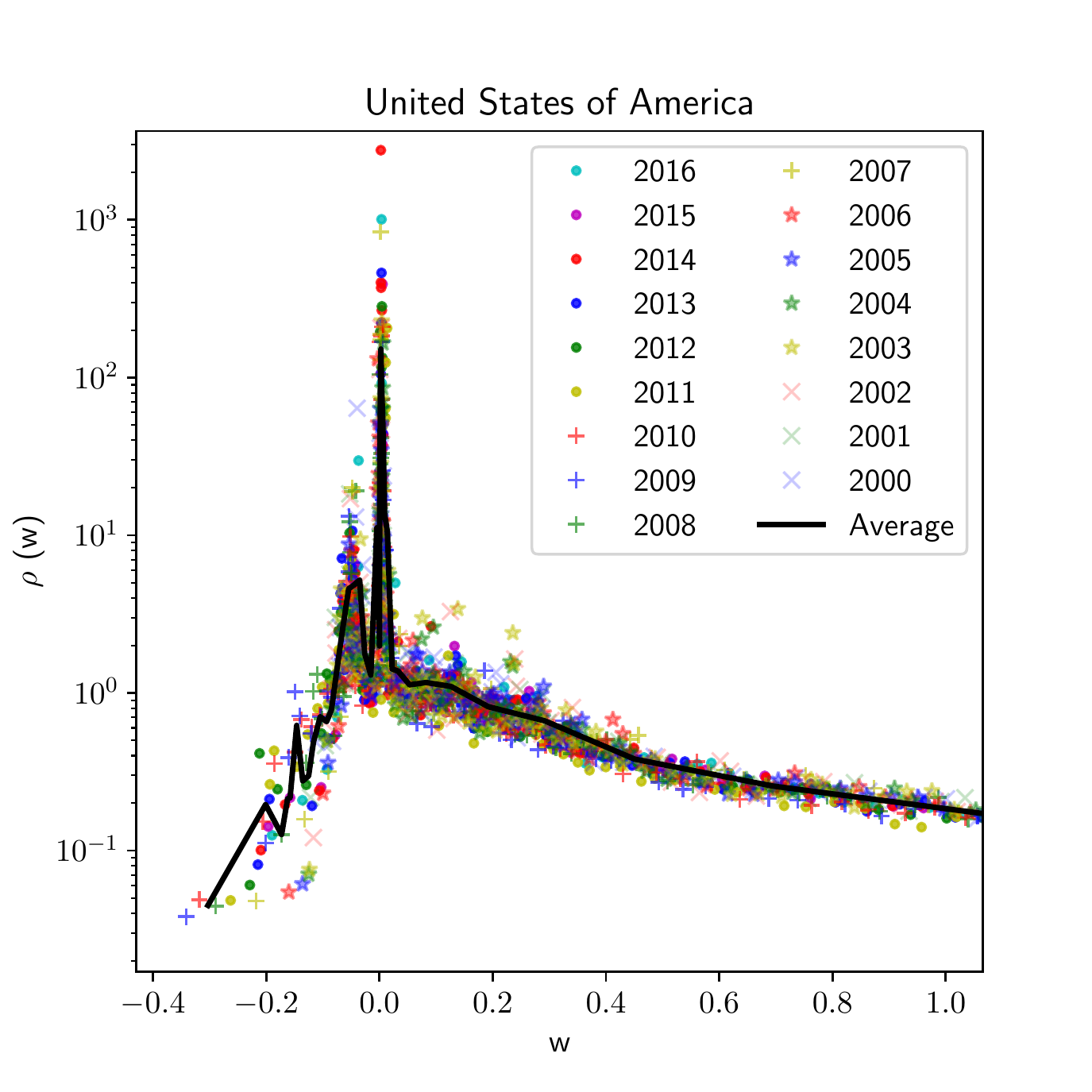}} 
    {\includegraphics[width=0.4\textwidth]{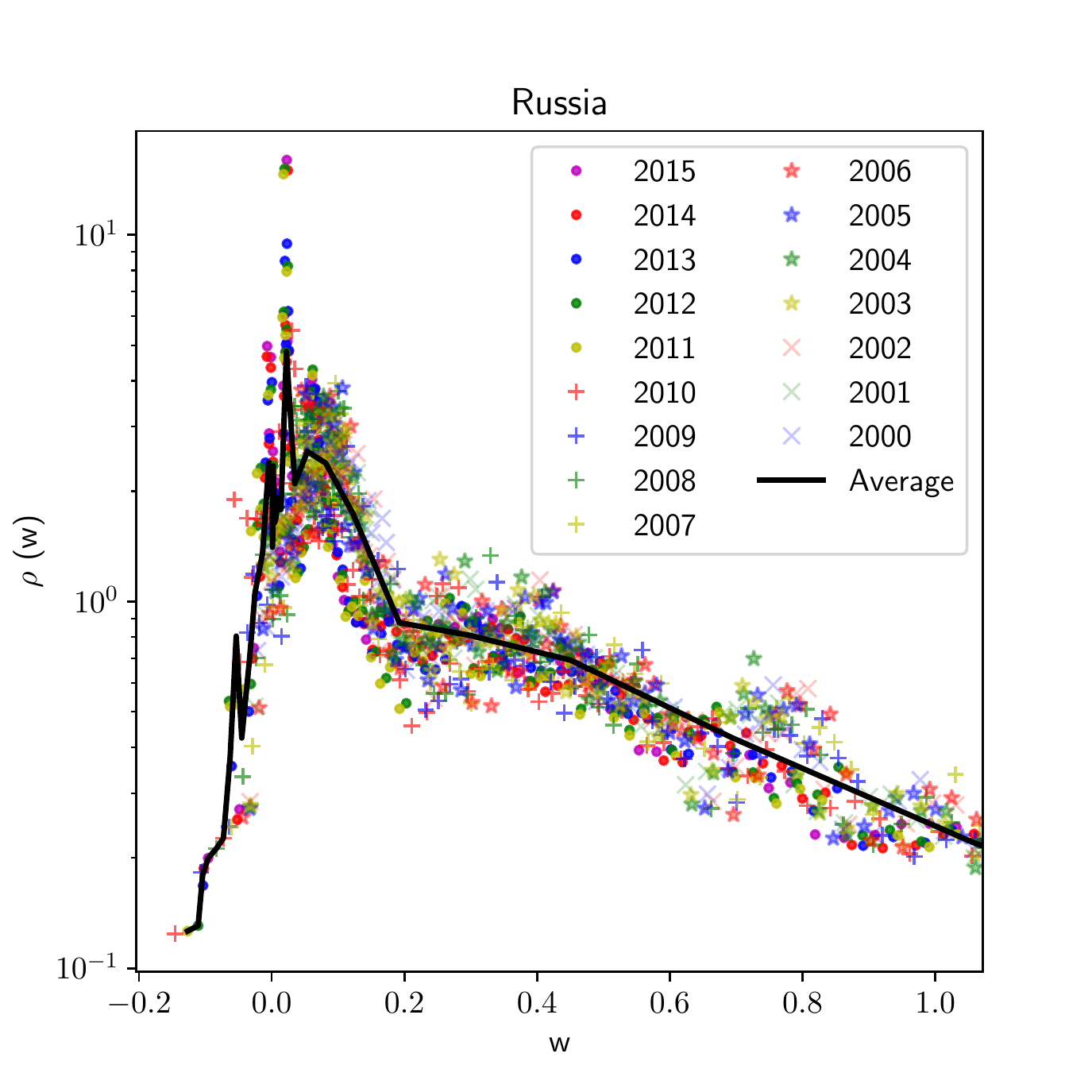}}     
\caption{Probability density function of the  normalized (rescaled) wealth, 
$w=W/\langle W \rangle$, for the negative (debts) and small wealth region. 
Results for USA, and Russia for the years indicated in the legends. The continuous 
black curve illustrates the average of the density functions for the considered years.}
    \label{collapse_neg}
\end{figure}

Due to the nature of wealth, which may be considered as the total sum of valuable 
possessions of an individual or a household,  quantifying its value is a complex task. 
Unlike income, there are no simple proxies that would offer the possibility of 
constructing an exhaustive dataset for the wealth distribution in a given 
geographic  region.  The methods used nowadays were already described in the last 
century \cite{atkinson}, and reconsidered in the recent years \cite{WIDBOOK}. 
Our wealth data are obtained from the World Inequality Database \cite{WIDdata}. 
These percentile datasets were derived from the National Accounts, Survey data, 
Tax data and Rich Lists
using complex methods, detailed in a working paper \cite{WIDBOOK}. 
First, we extracted data for USA and Russia (two economies with very different history) 
for several consecutive years.   
The probability density function (PDF) of the normalized wealth was computed for each 
country in each year, i.e. wealth was always normalized to the average wealth, 
$\langle W \rangle$, for the given year ($w=W/\langle W \rangle$). 
Using this method the wealth distributions for each country in different years collapsed 
on a master curve, as it is illustrated in Figures \ref{collapse_poz} and \ref{collapse_neg}. 
On Figure \ref{collapse_poz} we use log-log scales and plot the $W>0$ part of the 
density function. On Figure \ref{collapse_neg} we consider log-normal scales  and 
plot the part for negative (debts) and small wealth values.  
Taking into account the collapse visible in Figures \ref{collapse_poz} and 
\ref{collapse_neg} it becomes possible to derive  the averaged density function 
for each country in part, plotted on the respective graphs as  continuous black lines. 
 
\begin{figure}[!h]
    \centering
   {\includegraphics[width=0.45\textwidth]{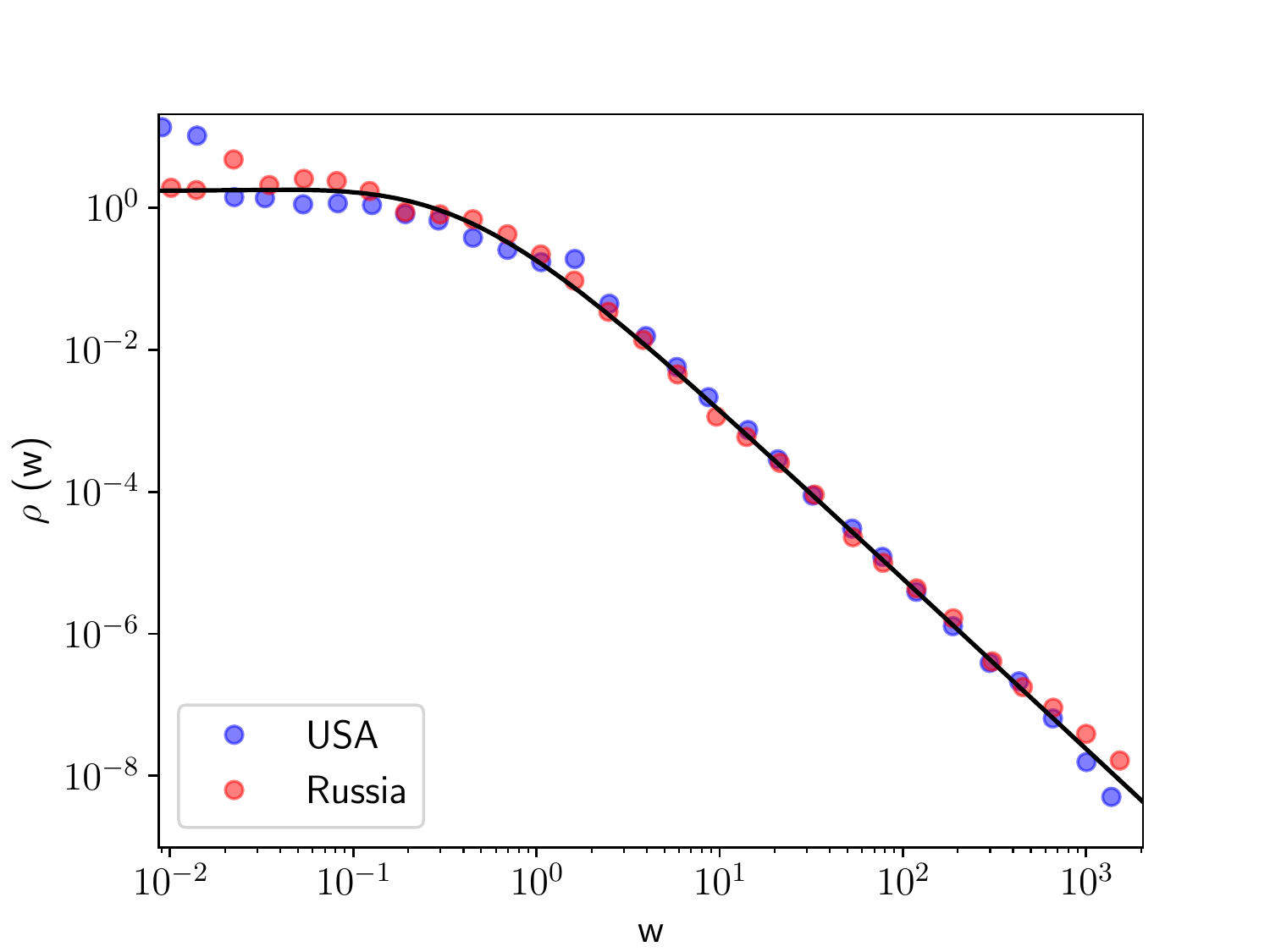}}
   {\includegraphics[width=0.45\textwidth]{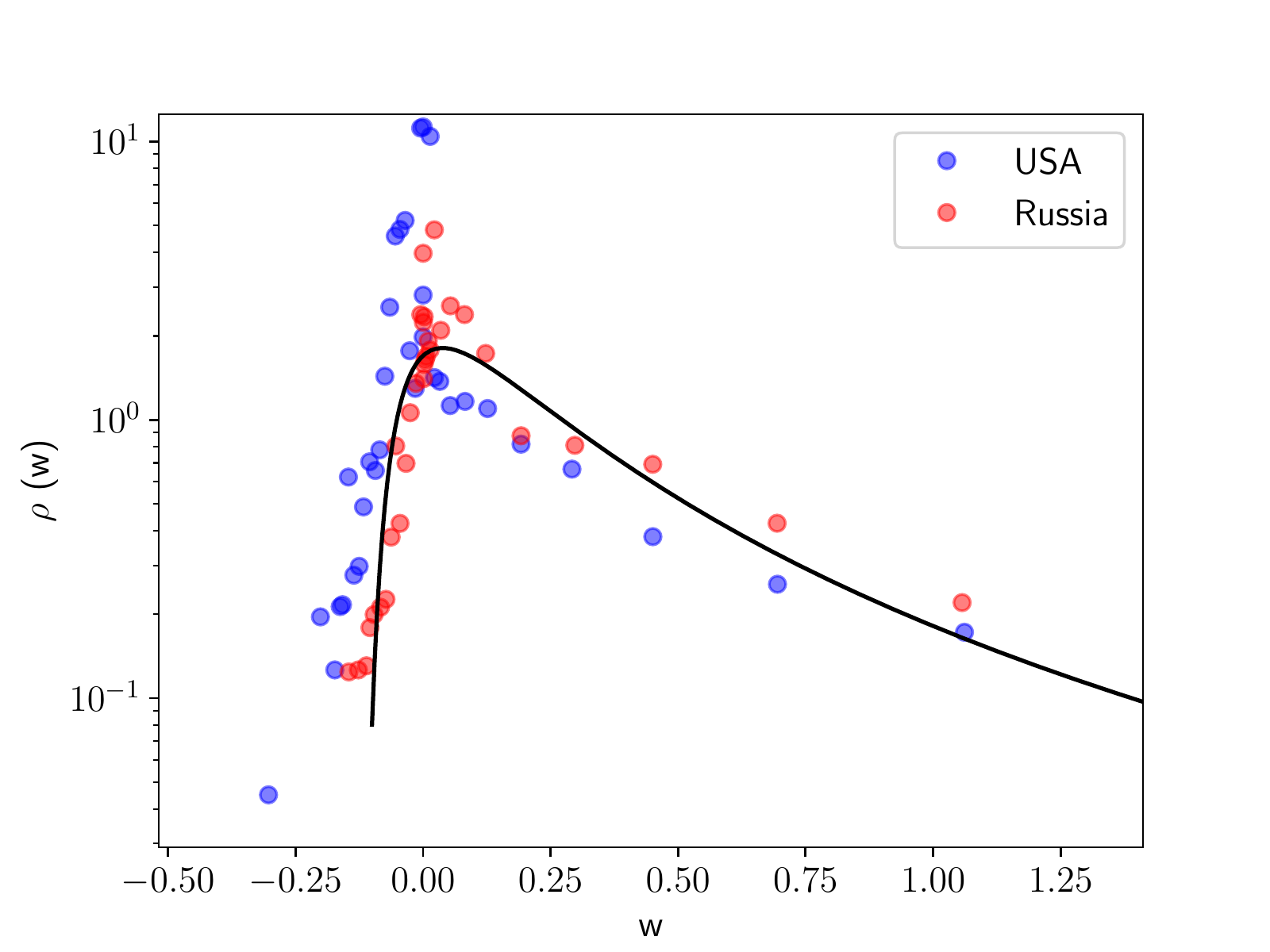}}
\caption{Probability density function of the  normalized (rescaled) wealth, 
$w=W/\langle W \rangle$. Qualitative comparison between data for USA and RUSSIA and 
the PDF given by our model, equation (\ref{df}). Disks with different colors 
(consult the legends of the figures) are the averaged results
for the studied years while the continuous thick line is the fit with the PDF 
given by eq. (\ref{df}), using the parameters: $k=1.4$,  $a=6.5$. Please be aware of 
the log-log scale for the figure on the left and the log-normal scale for the 
figures on the right hand-side.} 
    \label{comp}
\end{figure}

Interestingly, the trends both for USA and Russia are very similar and they can be 
compared in a better manner after plotting the average trends on the same curve. 
At this point one can attempt a fit with the density function (\ref{df}) obtained 
by our simple master equation approach with growth and reset terms. As it is illustrated 
in Figure \ref{comp}  the averaged PDF for the renormalized wealth is rather similar 
for the USA and Russia. We also learn from Figure \ref{comp} that one can obtain a 
qualitatively fair fit for the whole wealth interval using the parameters 
$k=1.4$,  $a=6.5$ in the PDF from equation (\ref{df}).

\begin{figure}[!h]
    \centering
   {\includegraphics[width=0.45\textwidth]{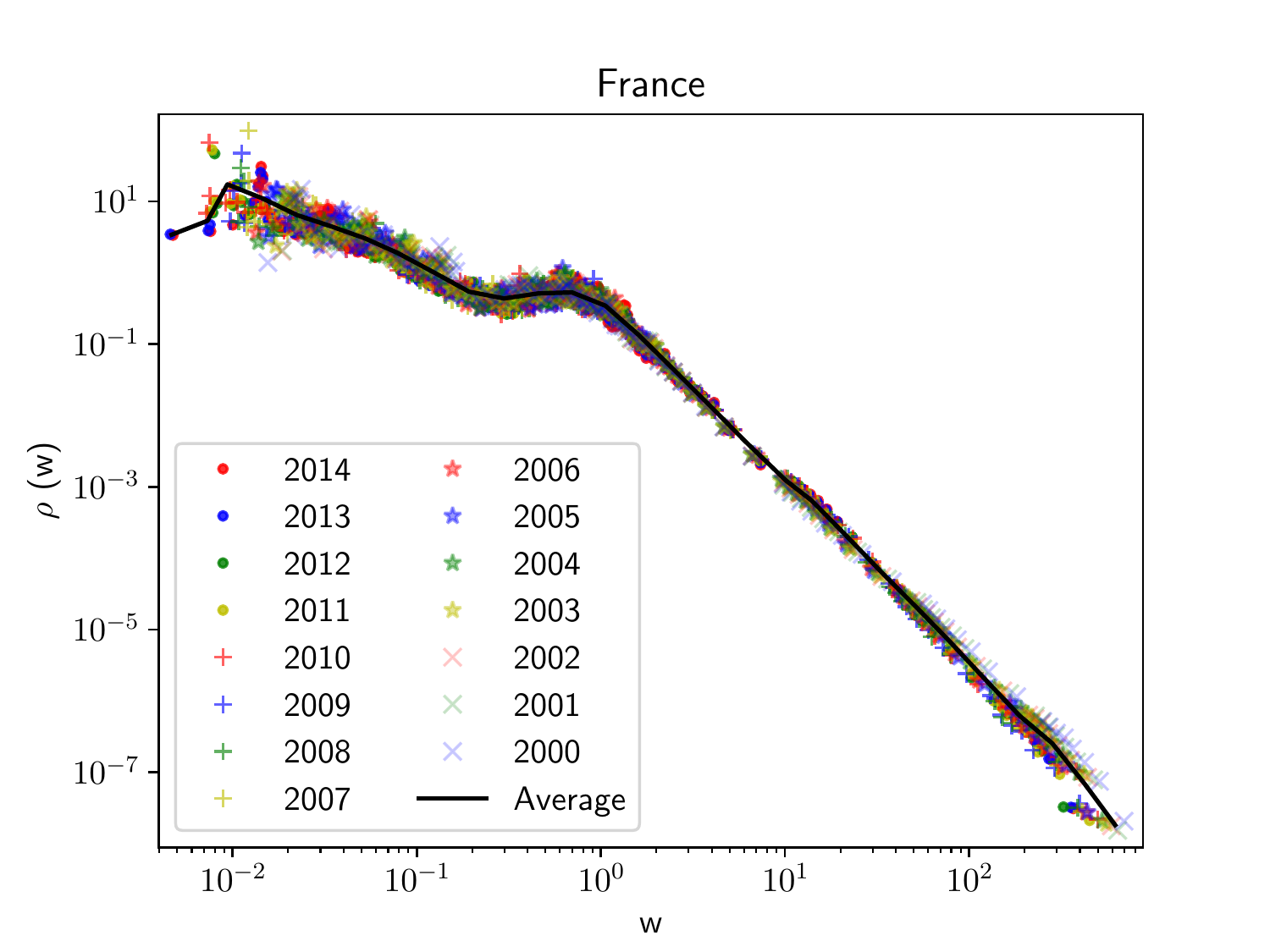}}
   {\includegraphics[width=0.45\textwidth]{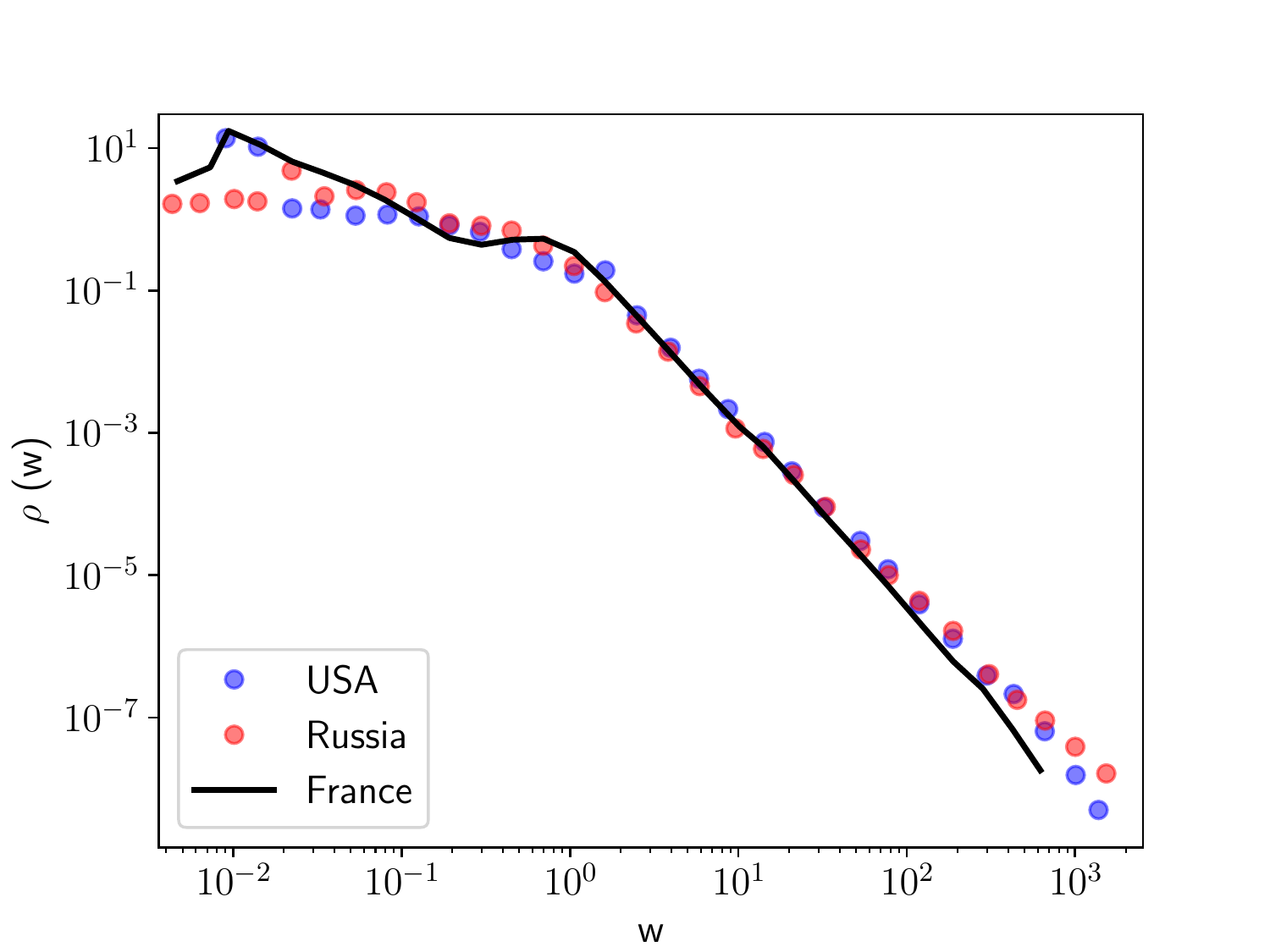}}
         \caption{Probability density function of the  normalized (rescaled) wealth, $W/\langle W \rangle$, for the $W>0$ region using log-log scales. The Figure on the left 
shows results for France for the years indicated indicated in the legends. The figure on the right hand-side compares the averaged PDF-s (average on all considered years) for 
USA, Russia and France.} 
    \label{comp-fra}
\end{figure}

According to this result one would assume thus an even a stronger universality for the PDF in the $W/\langle W \rangle$ variable. Not only the PDF for different years collapse, but Figure \ref{comp} suggests that also the PDF for different  countries might collapse on a universal trend. Performing an analysis on wealth distributions for other countries as well, we learn however that this is not the case. We can take for example the case of France, and extract the PDF of wealth distribution for several years from \cite{WIDdata}. Data for different years collapse again as it is indicated in  the left hand-side panel from Figure \ref{comp-fra}. (In case of France the data does not have 
information on negative wealth values, we used thus only the log-log plot for $W>0$. ) Plotting together the averaged PDF with the ones obtained for USA and Russia suggests that the fit parameters in this case should vary since the scaling in the limit of high wealth values is obviously different. In consequence, the collapse for the PDF in case of Russia and USA seems to be only a simple coincidence.

\section{Discussion}

\begin{figure}[!h]
    \centering
   {\includegraphics[width=0.45\textwidth]{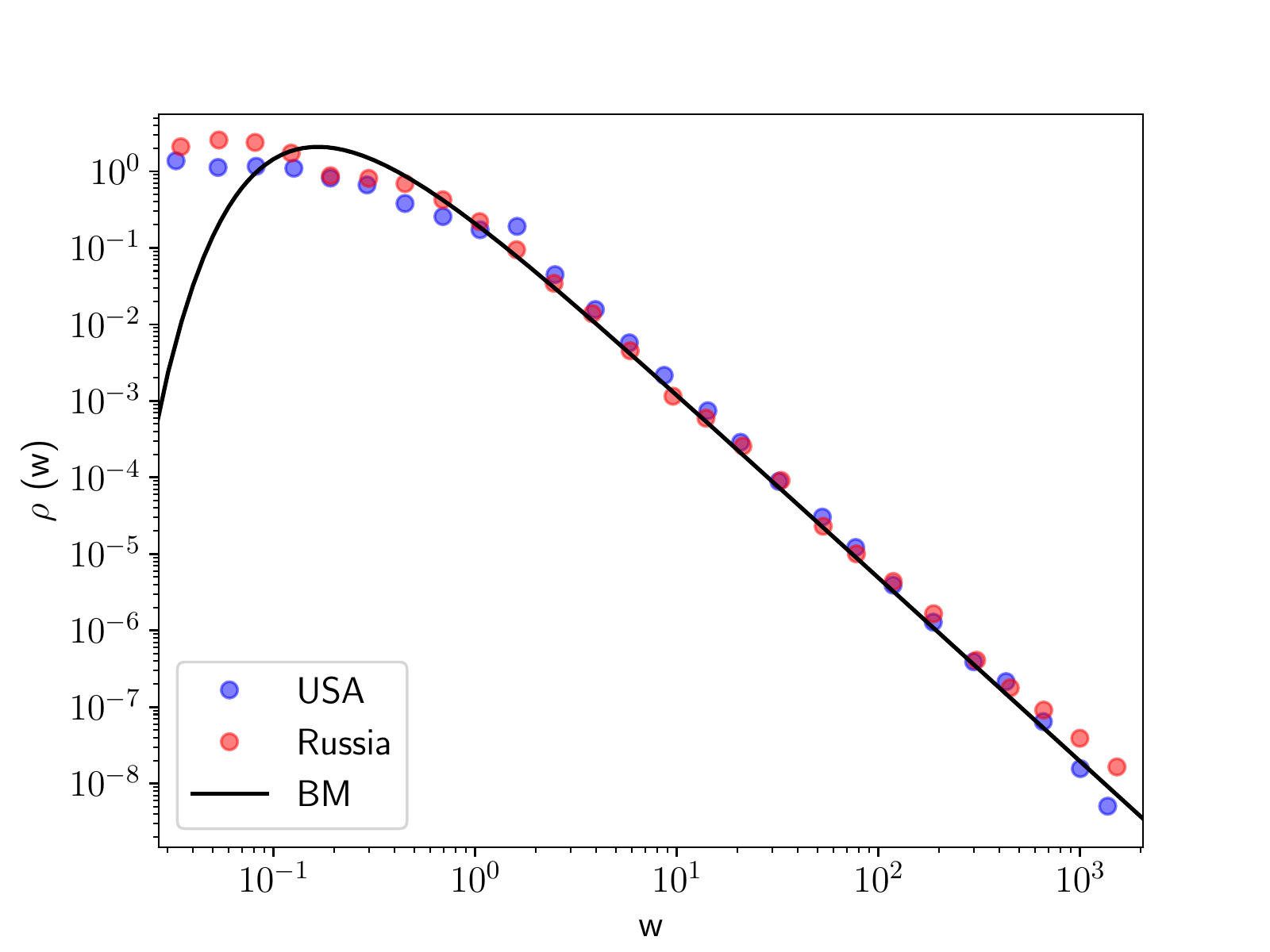}}
   {\includegraphics[width=0.45\textwidth]{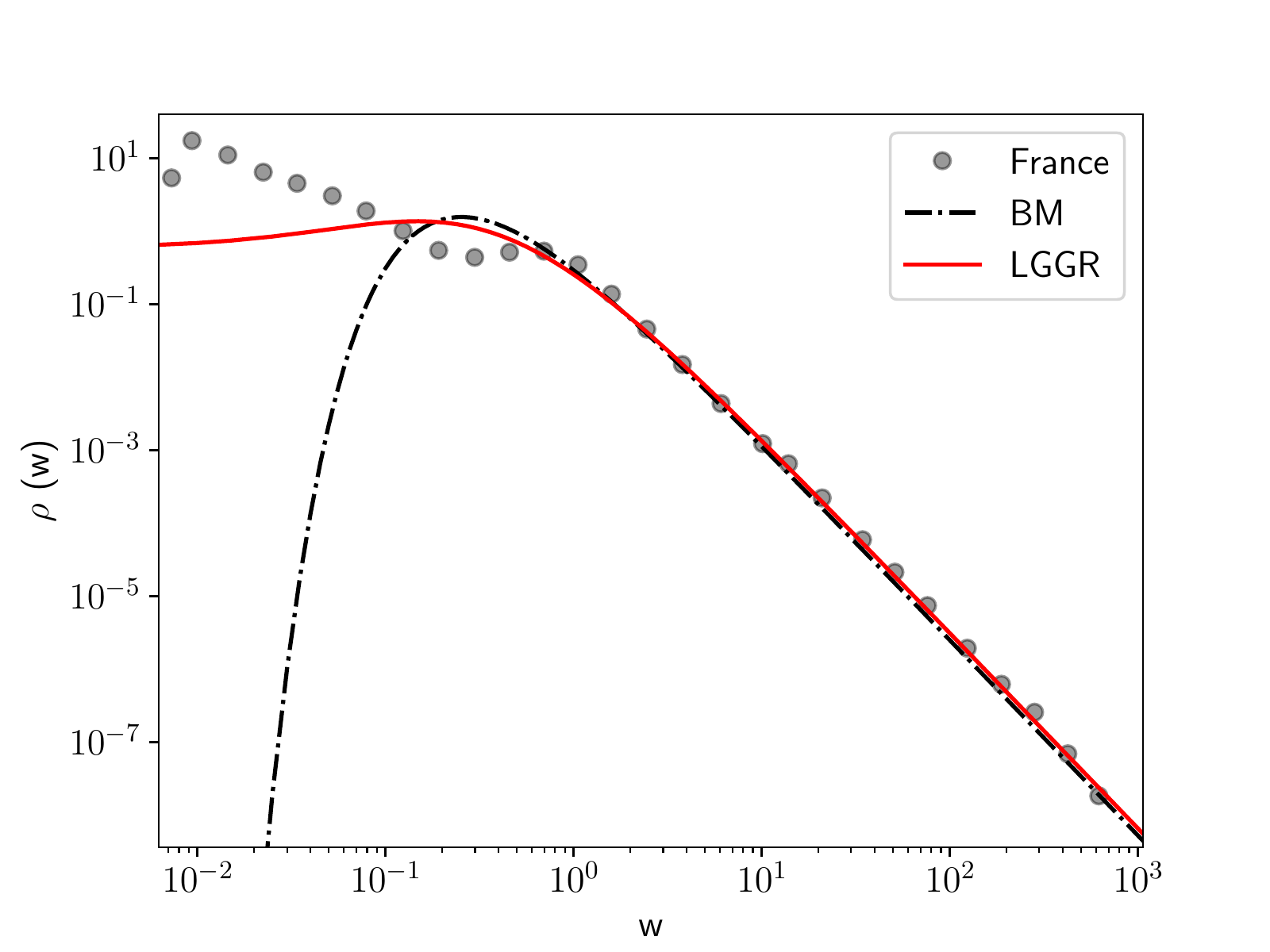}}     
\caption{Best fit for the data on the PDF of wealth distribution for USA, 
Russia, and France in the $W>0$ limit using the model of Bouchaud and Mezard (\ref{pdfBM}). 
On left hand-side panel we illustrate the fit with equation (\ref{pdfBM}) using  $g=0.4$. 
On the graph from the right hand-side we compare the best fits with the PDF's from 
equation (\ref{df}) to (\ref{pdfBM}). 
The corresponding parameters are: $k=1.68$, $a=7$ and $g=0.68$. } 
    \label{comp-BM}
\end{figure}

Comparison with collected data shows that the PDF given by equation (\ref{df}), 
obtained from a simple mean-field type growth and reset master equation, 
offers a good description for wealth inequalities in modern societies. 
Our results are also in agreement with the form of the PDF suggested by the 
alternative approach of Bouchaud and Mezard, considering a mean-field type exchange 
\cite{Bouchaud}. The advantage of the present model is however, that it allows to 
consider negative wealths (debts), too. Seemingly the two free parameters in the PDF 
from (\ref{df}), allows for an improved fitting even in the $W>0$ limit. 
In comparison with the fit given for $W>0$ in Figure \ref{collapse_poz}, 
on the Figure \ref{comp-BM} we illustrate the best fit 
($g=0.4$, leading to the same scaling law) that one obtains with the PDF  
from equation (\ref{pdfBM}).  Evidently our two-parameter LGGR model gives an 
improvement in the small wealth limit. This should not be a surprise, 
since the PDF from (\ref{pdfBM}) has only one fit parameter. 
The same observation is true if one considers the data for France. 
As it is visible in the right hand-side panel of Figure 
\ref{comp-BM} the best fit with equation (\ref{df}), 
(fit parameters: $k=1.68$ and $a=7$) is better for the small wealth limit 
when compared to the fit with equation (\ref{pdfBM}), (fit parameter: $g=0.68$). 
The scaling exponents in these cases coincide: $-2.68$.

The approach based on the growth and reset master equation is based on two 
hypotheses, formulated on the growth and reset rates, expressed mathematically by 
equations (\ref{rates}). 
It would be therefore in order to discuss here also the appropriateness of the rates 
used in our model. Concerning the chosen form for the growth rates we mentioned that 
it is in agreement with the generally accepted preferential growth hypothesis 
(Matthew's  principle). According to this, the wealth of individuals should grow with 
a speed that is proportional to their wealth values. 
This inevitably leads to an exponential increase, interrupted stochastically 
by the reset process. 

\begin{figure}[!h]
    \centering
\includegraphics[width=0.45\textwidth]{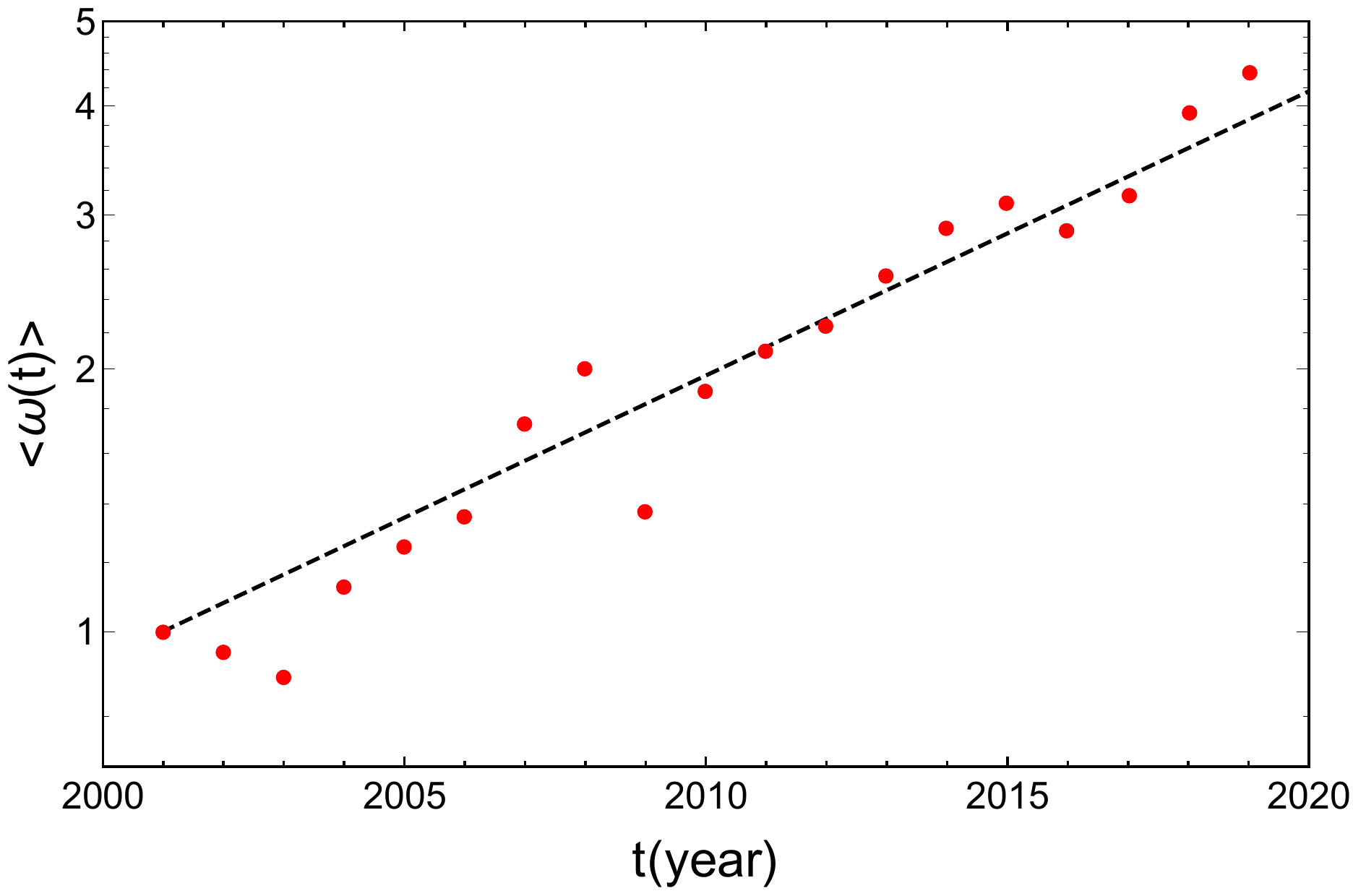}
\caption{Exponentially growing trend of the averaged rescaled wealth, $\omega(t)$, 
of the richest people in the World between 2001 and 2020. Please note that the 
vertical axis is logarithmic.} 
    \label{rich-growth}
\end{figure}

In order to support this hypothesis we investigated the growth of the wealth for 
the richest people in the world. We extracted from the Forbes database the  
15 leading persons  who were constantly in the top-list between 2001 and 2019. 
For each of them we followed their wealth $W_i(t)$ in each year, $t$, relative to the 
wealth from  2001, $\omega_i(t)=W_i(t)/W_i(2001)$, and then studied the average increase 
$\omega(t)=\langle \omega_i(t) \rangle_i$ as a function of time. The result for
$\omega(t)$ is plotted on Figure \ref{rich-growth} with log-normal scales. 
The apparently linear trend from Figure \ref{rich-growth} is in agreement with 
an exponentially trend: $\omega(t)\approx \exp[-0.075\:(t-2001)]$. 

\begin{figure}[!h]
    \centering
\includegraphics[width=0.45\textwidth]{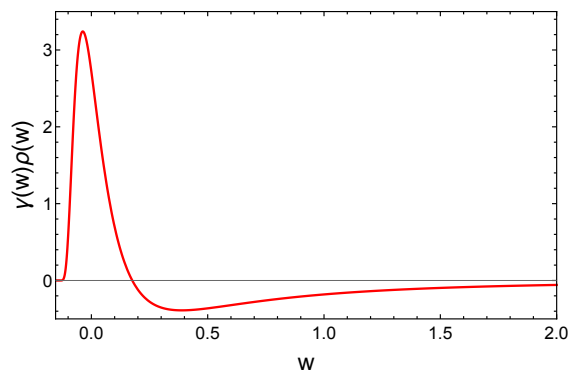}
\caption{Variation per unit time $dn(w)/dt\propto - \gamma(w)\, \rho(w)$ of the fraction 
of population with wealth around $w$ in a unit wealth interval ($n(w)=N(w,w+dw)/dw$) solely due to the smart reset process. 
The used model parameters are the ones given for the fit to the real world density functions 
from Figure \ref{comp} ($k=1.4$ and $a=6.5$). } 
    \label{reset-var}
\end{figure}

Concerning the reset rate, unfortunately we do not have any direct information 
supporting the used kernel function. It is possible however to show what are the 
consequences of this reset rate in view of the fitted PDF for USA and Russia. 
Using the fit parameters from Figure \ref{comp} ( $k=1.4$ and $a=6.5$), we determine 
the variations per unit time of the fraction of population in a unit wealth interval 
($n(w)=N(w,w+dw)/dw$) due to the smart reset process:
\begin{equation}
\frac{d\, n(w)}{dt}\propto - \gamma(w)\,  \rho(w)= -\left( k-\frac{(a+1)(k-1)}{1+a\,w} \right) \frac{a\, (a+1)^k\, (k-1)^k}{\Gamma(k)}  e^{ -\frac{(a+1)(k-1)}{1+a \, w}} \, (1+ a\, w)^{-1-k}\end{equation}
Results in such sense are plotted in Figure \ref{reset-var}. 
From here we learn that the majority of people start their dynamics in accumulated wealth 
in the region of small and negative  wealth values, well below the average wealth in the 
society ($w<0.2$) and leave the statistics with positive wealth. 
This is in nice agreement with our everyday-life experience. 

\section{Conclusion}

Following the modeling paradigm offered by the mean-field description of a 
stochastic growth and reset process we proposed a simple analytical model for 
wealth distribution in modern societies. The challenge we faced was to derive a 
compact mathematical formula for describing the density function of wealth distribution 
for all wealth categories, including the negative part (debts) as well.  
We used growth and reset rates similar with the ones used for modeling income 
distribution \cite{ZNeda1}.  The preferential growth rate is supported by wealth 
dynamics data of the world's leading billionaires. The smart reset rate is negative 
for the debtor and low wealth part of the society and becomes positive for middle and 
high wealth values. This is in agreement with our everyday observation that 
young  people usually start their life with debts or low wealth values and 
leave the society at an older age in the higher wealth categories. 
Moreover, bad transactions in the wealthier part of the society could reset the 
wealth values to negative or low values. 
Following this view we consider that the rates used by us are appropriate for 
approaching in a mean-field manner the dynamics of wealth in a society. 
The minor modifications in the growth and reset rates relative to the ones used 
for targeting the income distributions, allowed the extension to negative wealth values. 
As a consequence of these modifications we find that the shape of the stationary 
distribution function becomes different from the one obtained for the income. 
In \cite{ZNeda1} we argued that for income  a Beta Prime distribution function offers
a good description of the collected data. Here instead we find that a slightly modified 
version of the distribution function proposed by Bouchaud and Mezard \cite{Bouchaud} 
works most properly. The formula proposed in equation 
(\ref{df}) presents the Tsallis-Pareto type tail, and it gives a good fit for the data 
on the whole wealth interval. As a step forward relative to many earlier attempts for 
modeling wealth, this PDF describes in an acceptable manner the negative wealth limit, 
too. It is also important to note  that we found striking similarities between the 
wealth distributions in the USA and Russia. It is surprising in the view of their 
very different economic history. Similar to the income distribution, 
a rescaling by the average, the PDF-s of wealth for different years 
coincide on a master curve. 

\section*{Acknowledgements}
Work supported by the Romanian UEFISCDI  PN-III-P4-ID-PCE-2020-0647 research grant
and the Hungarian National Research Fund OTKA K 123815.

\section*{References}




\end{document}